\documentclass[a4paper,11pt]{article}
\usepackage{pos}
\usepackage{natbib}
\usepackage{floatrow}
\usepackage{wrapfig}
\title{Cosmic rays as a feedback agent in primordial galactic ecosystems}

\author*[a,b,c]{Ellis R. Owen}

\affiliation[a]{Department of Earth and Space Science, Graduate School of Science, \\Osaka University, Toyonaka, Osaka 560-0043, Japan}

\affiliation[b]{Institute of Astronomy, National Tsing Hua University, Hsinchu 30013, Taiwan (ROC)}

\affiliation[c]{Center for Informatics and Computation in Astronomy, National Tsing Hua University, \\ Hsinchu 30013, Taiwan (ROC)}

\emailAdd{erowen@astro-osaka.jp}

\abstract{High-redshift primordial galaxies have recently been found with evolved stellar populations and complex star-formation histories reaching back to 250 Myr after the Big Bang. Their intense bursts of star-formation appear to be interspersed with sustained periods of strong quenching, however the processes underlying this evolutionary behaviour remain unclear. Unlike later epochs, galaxies in the early Universe are not located in large associations like clusters. Instead, they co-evolve with their developing circumgalactic halo as relatively isolated ecosystems. Thus, the mechanisms that could bring about the downfall of their star-formation are presumably intrinsic, and feedback processes associated with their intense starburst episodes likely play an important role. Cosmic rays are a viable agent to deliver this feedback, and could account for the star-formation histories inferred for these systems. The cosmic ray impact on galaxies may be investigated using the wealth of multi-wavelength
data soon to be obtained with the armada of new and upcoming facilities. Complementary approaches to probe their action across the electromagnetic spectrum can be arranged into a distance ladder of cosmic ray feedback signatures. With a clear understanding of how cosmic ray activity in primordial systems can be traced, it will be possible to extend this ladder to high redshifts and map-out the role played by cosmic rays in shaping galaxy evolution over cosmic time.}

\FullConference{%
  *** 27th European Cosmic Ray Symposium - ECRS ***\\
  *** 25-29 July 2022 ***\\
  *** Nijmegen, the Netherlands ***
}

\begin{document}
\maketitle

\section{Introduction}

The formation and evolution of galaxies is an important topic, and it is essential to properly understand the processes that regulate their development.  Resolving the agents of feedback are a key aspect of this, particularly those at play within secular galaxies in the near and far Universe. Energetic supernova (SN) explosions have been well studied as a means to deliver thermal and mechanical feedback~\cite[e.g.][]{Oku2022ApJS}, and the effects of these mechanisms have been measured and quantified in detail~\cite[e.g.][]{Rosado1996AA}. Feedback dynamics in galaxies are therefore typically modeled mechanically, but this approach leaves important questions unanswered. What are the reasons for the decline in star formation after the cosmic noon? What causes the quenching of secular primordial galaxies fed by accretion of gas from the cosmic web? Why do abundance-matching relations between halo and stellar masses turn-over around the `golden mass' of $3\times 10^{11}$ $\text{M}_{\odot}$?

In addition to thermal and mechanical channels, there is a consensus that SN events also deliver substantial energy to their surroundings via cosmic rays (CRs), with SN shocks thought to be the dominant source of CRs in star-forming galaxies~\cite{Caprioli2012JCAP, BellAPh2013}. The influence of CRs is more challenging to observe than mechanical processes. They practically operate as a `hidden' pathway for energy and momentum feedback into their host galaxy. CRs are powerful energy carriers in galaxy ecosystems, and there is also evidence of their activity and engagement within the Milky Way and external galaxies (including signatures in $\gamma$-rays~\cite{Ackermann2012ApJ}, sub-millimeter~\cite{Indriolo2018ApJ} and radio~\cite[e.g.][]{Klein2018AA}). Based on energy budgets alone, CRs have clear potential as a hidden feedback agent in galaxy ecosystems~\cite[e.g.][]{Owen2018MNRAS} that is often missing in our current treatments of feedback dynamics. However, we are yet to settle on a complete picture of their local and global feedback impact within galaxies in different epochs. Constructing sensible procedures to investigate their effects in different host conditions is therefore crucial. 

\section{CRs in star-forming galaxies}

CRs are accelerated by violent, magnetised astrophysical environments. These include SN events and their remnants, where diffusive shock acceleration processes (e.g. Fermi acceleration~\citep{Fermi1949}) can energize charged particles to PeV energies~\cite{Fang2022PhRvL}. 
In a galaxy, the emergence of SN events is determined by the rate stars are being formed and the distribution of their masses (described by the initial mass function, IMF). If adopting a Salpeter IMF and stellar mass range between 1 and 50 ${\rm M}_{\odot}$, around 5\% of stars form core-collapse SN events, with their event rate tracking star-formation by $\mathcal{R}_{\rm SN} \approx 1.0 \;\! (\mathcal{R}_{\rm SF} / 160  \;\!  {\rm M}_{\odot} {\rm yr}^{-1}) \; {\rm yr}^{-1}$~\cite{Owen2018MNRAS}. These SNe transfer approximately 10\% of their mechanical energy to turbulence (with some variation depending on local conditions~\cite{Walch2015MNRAS}) and a similar amount to CRs~\citep[e.g.][]{Morlino2012AA}.

CR abundance is set by the star-formation rate in a galaxy when the accelerated CRs are contained by an interstellar magnetic field~\cite{Owen2018MNRAS}. In the local universe, star-forming galaxies provide evidence of this: they exhibit a close correlation between their $\gamma$-ray luminosity and markers of their star-formation activity~\cite{Ackermann2012ApJ}, where the high-energy emission is presumably a proxy for CR abundance. In the early Universe, some galaxies had much higher rates of star-formation than those in our local cosmic neighbourhood, with an increased incidence of strong starbursts among ancient galaxy populations~\cite{Stefanon2022ApJ}. It is likely that these periods of intensive star-formation produced an abundance of CRs, which would be a significant player in the structural evolution and assembly of galaxies in the early Universe~\cite{Owen2019MNRAS}.

\subsection{Secular galaxies at high redshift}

Star-formation in local galaxies is typically triggered by tidal processes associated with galaxy mergers and interactions. 
In a high-redshift setting, cold filamentary inflows can also be a major means of fuelling star formation, particularly in massive galaxies~\citep{Dekel2009Natur}. An accreting galaxy's star-formation rate is determined by how efficiently the gas supplied to it forms stars. There are several factors that can affect this, such as the hydrodynamic conditions in the vicinity of the galaxy, or feedback mechanisms that operate within the circumgalactic environment~\cite{Owen2019AA}. Recent observational studies have revealed young, secular primordial galaxies with multiple ages of convolved stellar populations. Their complex stellar formation histories (SFHs) extend back into the first 250 Myr after the Big Bang~\cite[e.g.][]{Hashimoto2018Natur}, and are characterized by intense bursts interspersed with periods of quenching. This evolutionary behavior appears to be endogenous in these relatively isolated ecosystems, with CRs serving as a plausible self-regulation delivery agent. The possibility of CRs playing a similar role in galaxies where starburst activity is triggered in a different manner may also be considered. 

\subsection{Post-starburst galaxies}
\label{sec:sec2p2}

Post-starburst galaxies (PSGs), also referred to as E+A galaxies, share similar evolutionary characteristics to primordial accretion-driven secular galaxies, including strong bursts and prolonged periods of quiescence. There is no guarantee that accretion from the cosmic web is the primary driver of star formation in PSGs. Instead, interactions and/or galaxy mergers may also be the triggers of their bursts~\cite{Chen2019MNRAS}. PSGs show evidence of having only recently shut-down their star-formation, and are undergoing a rapid transition in their star-formation properties from a burst phase to quiescence~\cite{Dressler1983ApJ}. They are typically identified by their distinctive spectral features. Specifically, they lack the optical emission properties expected from ongoing stellar formation, but exhibit strong Balmer absorption lines associated with young stellar populations. Moreover, they present evidence for a large old stellar population (identified by metallic absorption lines) indicating a long history of star-formation prior to their most recent burst. 

It is intriguing to consider the possible causes of the quenching of PSGs beyond their second resurgence of star formation. While their observed star formation rates are relatively low, CO-traced observations demonstrate that PSGs contain substantial molecular gas reservoirs~\citep{French2015ApJ}. Thus, the star-formation efficiency of these systems appears to be strongly suppressed. This implies that their evolution is being modified by feedback. To account for the properties and behaviour of these systems consistently, the feedback mechanism(s) at work seems to be progressive with a delayed-onset, and persists for a period after star-formation in the galaxy has already been shut-down. To determine whether CR processes could provide a viable mechanism consistent with these behaviours, it is necessary to assess how they operate in these systems.

\section{Cosmic rays as a feedback agent in galaxies}

\subsection{Energy deposition channels}

Protons of 1-10 GeV constitute the majority of the CR energy density in galaxies. These particles are therefore likely to be responsible for most CR energy feedback. Above a threshold of $\sim$ 0.3 GeV, the energy deposition channels available to CRs are primarily mediated through their hadronic interactions with interstellar gas nuclei. Proton-proton (pp) interactions produce secondary CR electrons and pion-decay $\gamma$-rays. $\gamma$-rays thus form a signature of CR activity in starburst galaxies~\cite{Ackermann2012ApJ, Owen2022MNRAS}, and coincide with the deposition of energy via the thermalization of secondary electrons. 

\begin{figure*}
    \centering
    \includegraphics[width=0.85\textwidth]{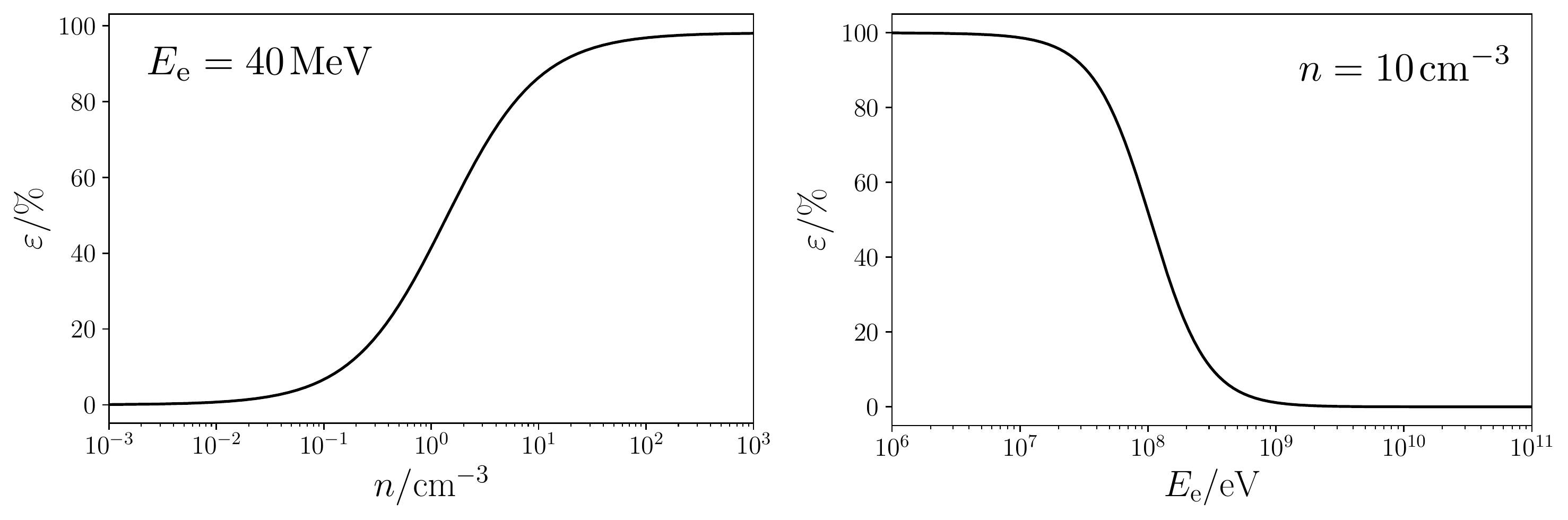}   
        \caption{Fraction of secondary CR energy available for thermalization into the ambient ISM. CR thermalization efficiency is strongly influenced by local conditions (e.g. density; left panel) and the competition with loss inefficiencies (e.g. radiative cooling), which is energy-dependent. Figure adapted from~\cite{Owen2018MNRAS}.}
\label{fig:heating_efficiency}
\end{figure*}

The overall CR energy feedback efficiency is determined by the thermalization mechanism appropriate for the secondary electrons. Thermalization proceeds via Coulomb scattering. Inefficiencies are then caused by competing radiative losses, and depend strongly on local physical conditions (gas density, ionization fraction, magnetic field strength, ambient radiation fields). Consequently, the average CR heating effect may be significant across a galaxy~\cite{Owen2018MNRAS}, but it can vary substantially within a complex, multi-phase interstellar medium (ISM; see Figure~\ref{fig:heating_efficiency}), with the strongest effects felt in regions with high density and ionization fraction.

\subsection{Magnetic containment}

CRs propagate at the speed of light but are deflected by ambient magnetic fields. In a magnetized ISM, these deflections are local, but they accumulate practically continuously along CR propagation paths. Macroscopically, this slows their effective speed and establishes a diffusive regime. CR escape from a galaxy then becomes hindered, and their abundance is set by the steady-state balance of their acceleration against diffusive leakage into circumgalactic space (determined by the ISM magnetic field strength). The resulting containment effect can enhance the CR density in a galaxy by 5-6 orders of magnitude compared to a ballistic CR streaming scenario~\citep[see][]{Owen2018MNRAS}, refocusing their feedback power back into their host.

\begin{figure*}
    \centering
    \includegraphics[width=0.65\textwidth]{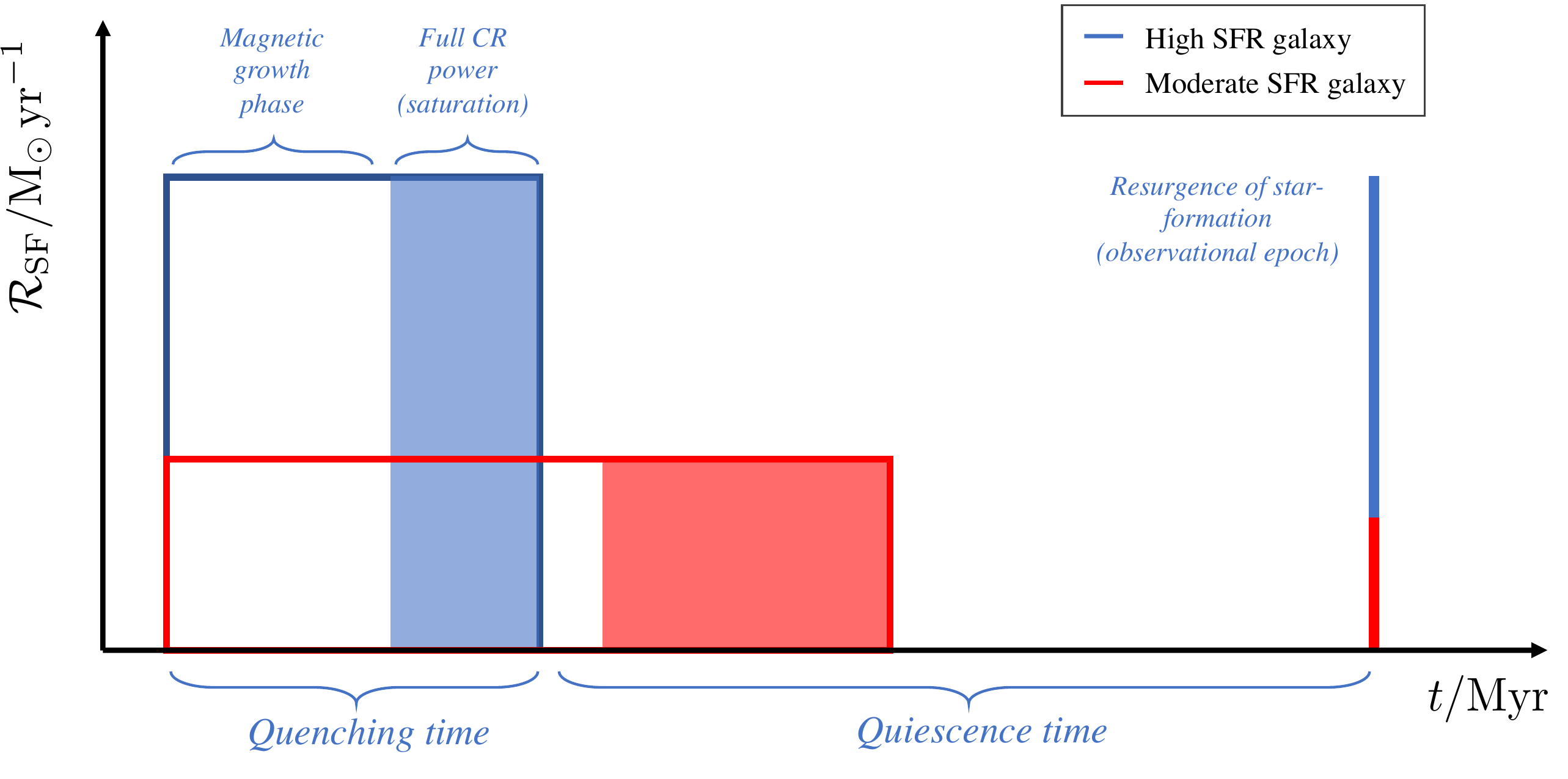}   
        \caption{Schematic of the evolving star-formation rate $\mathcal{R}_{\rm SF}$ of a galaxy over time, when regulated by CRs. A highly star-forming galaxy (shown in blue) develops a saturated magnetic field relatively quickly. Containment of CRs rapidly brings about the downfall of star-formation in the galaxy, before it later undergoes a resurgence. A lower $\mathcal{R}_{\rm SF}$ galaxy (red line) takes longer to amplify its magnetic field, longer to generate a sufficient abundance of CRs to halt star-formation and sees a quicker resurgence. 
}
\label{fig:sfhistory}
\end{figure*}

As a galaxy's magnetic field evolves, its ability to contain CRs changes. As early as $z\sim 1$, there is evidence of coherent $\mu$G strength fields in galaxies~\cite{Bernet2008Natur}. More recent studies further indicate that  strong, coherent magnetic fields were present in the ISM of quasar host galaxies by $z\sim 6$~\cite{Lee2022arXiv}. To account for this rapid magnetization, it has been proposed that galaxies' magnetic field development may be associated with their star-formation~\cite{Schober2013AA}. In this scenario, a turbulent dynamo mechanism can be invoked where magnetic amplification is caused by the turbulence injected by SN events during a galactic starburst episode. In this picture, it takes some time for the magnetic field of a galaxy to grow. This translates into a delay before the full power of CR feedback is then felt~\cite{Owen2018MNRAS}. CR heating continues beyond the end of the starburst phase, until the magnetic field dissipates sufficiently to weaken its containment effect, or until enough CRs leak away.

\subsection{Evolutionary properties of CR-regulated galaxies}
\label{sec:timescales_info}

A galaxy's magnetic field must saturate before maximum CR feedback power can be delivered. This delays the full effects of CR heating until some time after the onset of star-formation. In more intensely star-forming systems, feedback impacts are stronger, and the full impact is felt by the galaxy more quickly. Conversely, in more quiescent systems, feedback is weaker and the amplification of the magnetic field takes more time. Thus, with stronger feedback delivery: (i) the timescale of a starburst (the delay before sufficient magnetic containment is achieved) is shorter, and (ii) the time until a resurgence of star-formation is longer, if relying on the cooling of gas already held in the galaxy's reservoir or the reinstatement of accretion flows~\cite{Owen2019AA}. CR feedback thus yields a hallmark signature in the SFHs of both PSGs and primordial secular accretion-driven protogalaxies, where the duration of the post-starburst quenched epoch is proportional to star-formation rate inferred during a the burst phase $\mathcal{R}_{\rm SF}^{\star}$, and the duration of the burst is inversely proportional to $\mathcal{R}_{\rm SF}^{\star}$ (see Figure~\ref{fig:sfhistory}). 

\section{Discerning CR feedback}

\subsection{A distance ladder of cosmic ray feedback signatures}

There may never be a clear, accessible `smoking gun' for CR feedback applicable to all galaxies. However, a robust framework may still be constructed to identify their activity at different epochs, akin to the cosmic distance ladder. In the local Universe, resolved $\gamma$-rays are an effective tool to reveal the action of CRs in individual galaxies~\citep{Ackermann2012ApJ}. During the high-noon of cosmic star-formation, unresolved $\gamma$-ray backgrounds can instead be useful~\citep{Owen2022MNRAS}, or astrochemical signatures of their ionization effects~\citep{Indriolo2018ApJ}. At even further distances, CR activity in PSGs and isolated high-redshift protogalaxies may be constrained by reconstruction of SFHs to allow estimation of quenching/quiescence timescales in individual or stacked galaxies (cf. section~\ref{sec:timescales_info}). This is possible using a combination of the forbidden [\ion{O}{iii}] emission line flux at 88$\mu$m to estimate the observed star-formation rate of a galaxy, and the spectral energy distribution (in particular, the Balmer break) to determine the age(s) of its stellar population(s). Suitable data to support such efforts is already becoming available with the Atacama Large Millimetre/Sub-Millimeter Array (ALMA) and the NIRSpec instrument on the \textit{James Webb Space Telescope} (\textit{JWST}).

\subsection{Degeneracies at the top of the ladder}

\begin{figure}
\floatbox[{\capbeside\thisfloatsetup{capbesideposition={right,bottom},capbesidewidth=6cm}}]{figure}[\FBwidth]
{\caption{A sample of PSGs for which sufficient information was available to estimate quenching and quiescent timescales. The negative correlation between these quantities is indicative of an endogenous feedback mechanism. Figure adapted from~\cite{Owen2019AA}. \vspace{0.5cm}}
\label{fig:timescales_example}}
{\includegraphics[width=8cm]{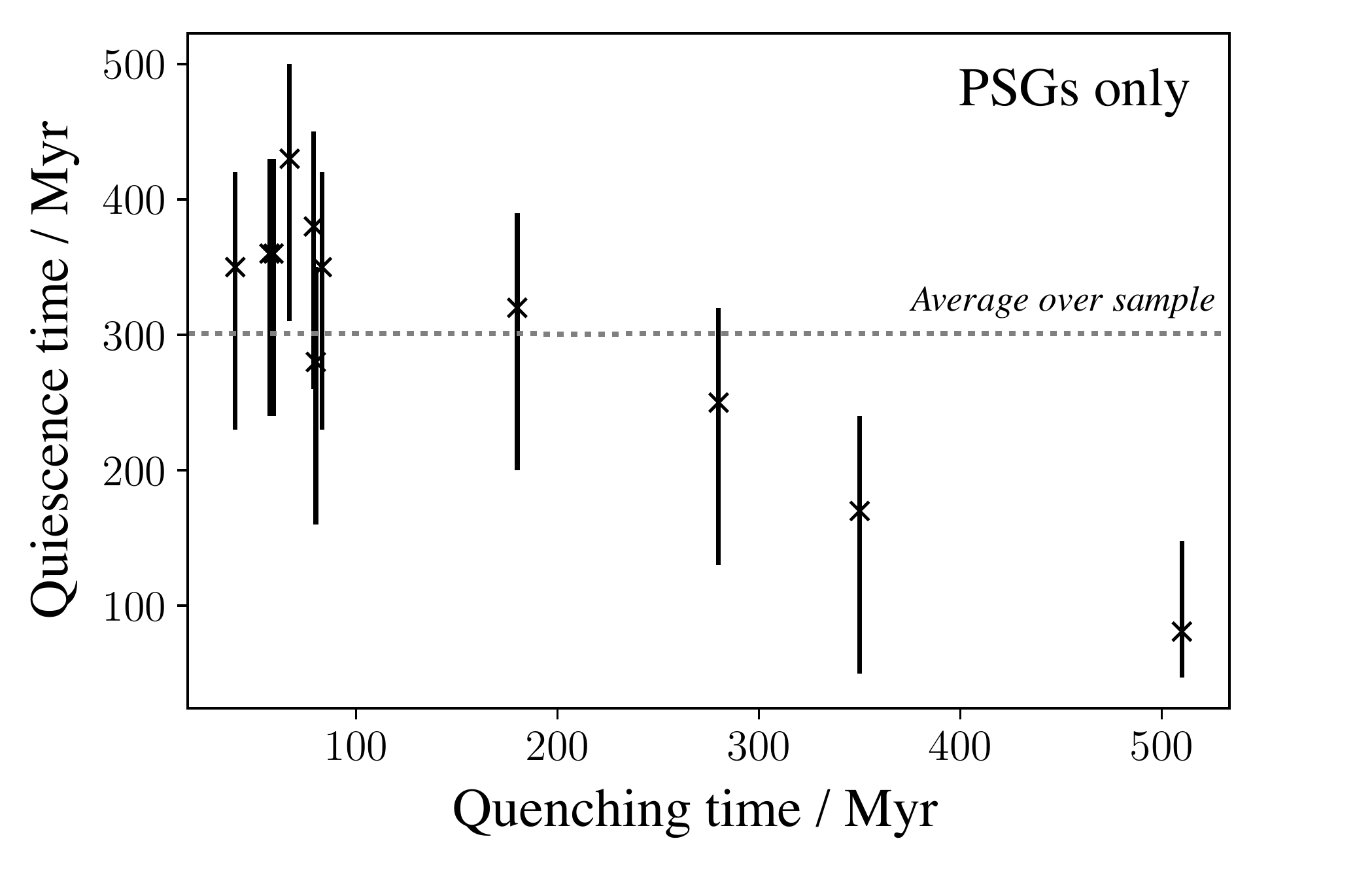}}
\vspace{-0.5cm}
\end{figure}

Differentiating CR feedback from other viable mechanisms is crucial in the high redshift context (i.e. at the `top' of the distance ladder). A population study can be useful in this regard: CR feedback is progressive and delayed. Moreover, the speed it takes hold in a galaxy and the extent it quenches star-formation is directly related to the star-formation rate. Across a population of galaxies, this endogenous behaviour is revealed by a clear correlation between proxies for feedback power and feedback impact, such as quenching speed (length of starburst episode) and quenched duration. A scenario where an endogenous process dominates the evolution of a galaxy population results in an inverse correlation between these quantities, with minimal scatter. This is demonstrated in Figure~\ref{fig:timescales_example}, where a clear inverse proportionality between quiescence and quenching time emerges. By ruling out other possibilities, a stronger indication of CR feedback activity can be obtained. For instance, the trends apparent in Figure~\ref{fig:timescales_example} would not arise under a more stochastic feedback mechanism, such as energetic hyper-nova events: feedback would instead be delivered suddenly and randomly instead of progressively, yielding scattered timescales without any clear intrinsic dependency on galaxy properties. Other mechanical feedback mechanisms are easier to rule-out in PSGs  - outflows that evacuate gas are unlikely to be active, since considerable gas reservoirs remain (see section~\ref{sec:sec2p2}). Additionally, although they can be present, AGN have been found not to play a major role in shaping PSG evolution~\citep{Lanz2022ApJ}.

\vspace{-0.2cm}
\section{Summary and opportunities}

\vspace{-0.2cm}
Understanding the role of hidden players in galaxy evolution remains an important topic - in particular identifying the agents of feedback that operate through means other than mechanical energy. CRs have clear potential as such an agent. With the broad range of tools we have at our disposal, we can now begin to map-out their activity using a distance ladder of cosmic ray feedback, together with the wealth of multi-wavelength information soon to be obtained with the armada of new and upcoming facilities. A particular challenge lies at the most distant end of this ladder, where high energy signatures are not available and chemical signatures are not accessible. Instead, the opportunity is now within reach to measure spectra and star-formation rates with ALMA and \textit{JWST} for individual lensed, distant starbursts and PSGs to obtain their SFHs. This opens-up the prospect of soon being able to identify a new signature of CR feedback in galaxies at high redshift, and unveiling their role in the structural evolution of the Universe. 

\vspace{-0.2cm}
\acknowledgments

\vspace{-0.2cm}
\noindent
This work was supported by the Japan Society for the Promotion of Science (JSPS) KAKENHI Grant Number JP22F22327. ERO also acknowledges support from a grant from the Ministry of Education of Taiwan (ROC) at the Center for Informatics and Computation in Astronomy, National Tsing Hua University, where some of this work was completed.



\bibliographystyle{ECRS}
\bibliography{references}

\end{document}